# New parameters of Non-commutativity in Quantum Mechanics

Corresponding Author

**Mostafa Ijavi**

Department of Energy Engineering & Physics, Amirkabir University of Technology, Tehran, Iran

*Corresponding author e-mail: m.ijavi67@gmail.com and m.ijavi@aut.ac.ir

**Abstract**

At this paper, it is considered to find a way for defining non-commutative spaces by ordinary commutative ones and vice versa. A novel parameter which has not been considered so far is represented. This parameter describes equivalent spaces. Also, we searched concepts of these new parameters with one problem. Noncommutativity in total space is important here because it could explain more concepts. As we knew SW method (Seiberg-Witten) explained noncommutativity so here, we showed that it was not suitable for some conditions.in the end we considered Hamiltonian of free particle in new noncommutativity and we found concepts of new parameters.

**1. INTRODUCTION**

The structure of space-time non-commutativity in Quantum field theory was considered originally by Snyder (1947), Heisenberg (1930), Pauli (1946) and Yang (1947). Theories in non-commutative space have been surveyed extensively (Douglas and Nekrasov 2001). Non-commutative spaces arise in some phenomena at various fields of physics such as Field theory, String theory, M-theory (Connes and Douglas and Schwarz 1998; Seiberg, Witten 1999; Carroll et al. 2001; Delduc et al 2008), Quantum gravity and condensed matter physics quantum Hall effect (Leonard 2001; Bellissard and Van Elst and Schulz-Baldes 1994) as well as topological insulators. Non-commutativity in quantum field theory can be achieved in two different ways: via Moyal -product on the space of ordinary functions, or defining the Field theory on a coordinate operator space which is intrinsically non-commutative (Douglas and Nekrasov 2001; Chaichian and Demichev and Presnajder 2000).

Assumptions (Daszkiewicz and Walczyk 2008; Martinetti and 2005; Scholtz and Gouba and Hafver 2009; Bastos and Bertolami and Costa Dias and Nuno Prata 2008) representing that the space-time coordinates do not commute, neither in a canonical way

$$[\hat{x}_i,\ \hat{x}_j] = i\theta_{ij},\ [\hat{p}_i,\ \hat{p}_j] = 0,\ [\hat{x}_i,\ \hat{p}_j] = i\hbar\delta_{ij},\ [\hat{x}_i,\ t] = 0, \qquad i,j = 1,2,3,\dots,d \qquad (1)$$

Where $t$ is time, $\hat{p}_i$ is momentum in non commutative space, $\hat{x}_i$ is non commutative space coordinates, $\theta_{ij}$ is antisymmetric real constant ($d\times d$) matrix, $\delta_{ij}$ is the identity matrix and $d$ is dimension.

If the momentum is non-commutative, the equation 1 could be rewritten as below:

$$[\hat{x}_i,\ \hat{x}_j] = i\theta_{ij}, [\hat{p}_i,\ \hat{p}_j] = i\eta_{ij}, [\hat{x}_i,\ \hat{p}_j] = i\hbar\delta_{ij}, [\hat{x}_i,\ t] = 0, \qquad i,j = 1,2,3,\dots,d \qquad (2)$$

Equation 2 is non-commutativity in the total phase space where $\eta_{ij}$ is antisymmetric real constant ($N\times N$) matrix.

Theoretical predictions for specific non-commutative systems have been compared with experimental data leading to bounds on the non-commutative parameters (Chaichian 2000; Szabo 2003):

$$\theta \leq 4 \times 10^{-40} m^2, \qquad \eta \leq 1.76 \times 10^{-61} \frac{m^2}{s^2} \qquad (3)$$

It is demonstrated that by using the Seiberg-Witten (SW) maps (Seiberg and Witten 1999, Nuno CD, Maurice DG, Franz L, Joao NP 2010), it is possible to define the non-commutative space with the use of commutative space and it has been demonstrated that there is a linear relationship between non-commutative and commutative space.

In this paper, the Seiberg-Witten (SW) maps are modified and a novel model by using Heisenberg Algebra has been obtained for the relationship between the non-commutative and commutative spaces.

**2. Analysis of spatial non-commutativity for free particle**

In this part, the influences of spatial non-commutativity for free particle in quantum equations are going to be discussed and, its results on this matter is going to be checked. As we know, free particle Hamiltonian, (with the mass of $m_p$ and linear momentum of $p$) is $H = \frac{p^2}{2m_p}$.

There is no impact on free particle Hamiltonian if the space is non-commutative and momentum is commutative. Hence the amount of energy does not change in non-commutative space condition therefore, the thing is: Does non-commutativity effect on energy's Eigen value?

The response is NO and it is concluded that so why the calculations of non-commutativity are used while no new meaningful component is used in this issue as it is clear here.

This matter is declared on another sight. Assume the particle has an electrical disorder. It means that $H = \frac{p^2}{2m_p} + \varepsilon e x_3$ ($x_3$ is space dimension, $e$ is electric charge of the particle and $\varepsilon$ is uniform electric field on the particle). $\varepsilon e x_3$ is the electric potential. And now, if the space is considered as non-commutative, then the Hamiltonian is $\hat{H} = \frac{\hat{p}^2}{2\hat{m}_p} + \hat{\varepsilon}\hat{e}\hat{x}_i$. Here $\hat{H}$ is non-commutative space Hamiltonian, $\hat{\varepsilon}$ is field in non-commutative space, $\hat{e}$ is charge in non-commutative space, $\hat{m}_p$ is mass in non-commutative space and $\hat{x}_3$ is space dimension in non-commutative space.

Space noncommutativity effect on disorder Hamiltonian but did not affect simple one. Here, there are some questions: how can we demonstrate the non-commutativity before the disorder? Is the electrical field an entrance for it? Does the electrical disorder have some errors?

Before responding to these questions, another kind of disorders is mentioned which is a magnetic type of them. In fact, electrical field and magnetic field classify in the same, weak power group which are the standard model form of their own nature so, they have the same impact on the space. Assume that a free charged particle is located in a uniform weak electrical field which is named B. The Hamiltonian equation transforms to $H = \frac{\left(p_0 + \frac{e}{c}A\right)^2}{2m_p} = \frac{p^2}{2m_p}$, thus Hamiltonian's structure is not the same in commutative and non-commutative spaces. It means that the magnetic disorder can signify the spatial non-commutativity.

## 3. Investigation of spatial non-commutative

First of all, the simpler section which is spatial non-commutativity is used for getting to know the method of calculating in this paper then; main calculations are going to be studied such as non-commutativity in the whole space like momentum and space. By using of SW (Seiberg-Witten) (1999) map, the spatial non-commutativity equations are obtained (only space is non-commutative, time and momentum are commutative):

$$\hat{x} = x - \frac{\theta p}{2\hbar}, \qquad \hat{p} = p \tag{4}$$

But in the represented method, the equations are expressed in another way. Due to Heisenberg algebra's linearity, the equations below can be presented for commutative and non-commutative spaces:

$$\hat{x}_i = A_{ij}x_j + B_{ij}p_j, \qquad \hat{p}_i = C_{ij}x_j + D_{ij}p_j, \qquad i,j = 1,2,3,\dots,d \tag{5}$$

The Js are totals. $A_{ij}, B_{ij}, C_{ij}, D_{ij}$ are matrix coefficients for A, B, C, D matrixes. They are obtained by using equations (1). In fact, matrix form of the equations (5) is:

$$\hat{x} = Ax + BP, \qquad \hat{p} = Cx + Dp \tag{6}$$

By using equations (1) and (5):

$$A_{im}B_{jm} - A_{jm}B_{im} = \frac{\theta_{ij}}{\hbar}, \qquad C_{im}D_{jm} - C_{jm}D_{im} = 0, \qquad A_{im}D_{jm} - C_{jm}B_{im} = \delta_{ij} \tag{7a}$$

$$\theta_{ij} = -\theta_{ji}, \qquad \theta_{ii} = 0 \tag{7b}$$

Where $\theta_{ij}$ is symmetric

By knowing the matrixes

$$AB^T - BA^T = \frac{\theta}{\hbar}, \qquad CD^T - DC^T = 0, \qquad AD^T - BC^T = I_{d \times d}, \qquad \theta = -\theta^T \tag{8}$$

$A, B, C, D, \theta, I_{d \times d}$ are the matrixes with entries $A_{ij}, B_{ij}, C_{ij}, D_{ij}, \theta_{ij}, \delta_{ij}$, respectively. There are $4d^2$ ($d$ is dimension) parameters in equations (9), (10) and (11), with $\frac{3d^2 - d}{2}$ independent equations and $\frac{5d^2 + d}{2}$ dependent equations.

The matrixes $A$ and $D$ are assumed as unit matrixes. so

$$A = I_{d \times d}, \qquad D = I_{d \times d} \tag{9a}$$

$$B^T - B = \frac{\theta}{\alpha \hbar} \tag{9b}$$

$$C = C^T \tag{9c}$$

$$BC^T = 0 \tag{9d}$$

Here $\alpha$ and $\beta$ are arbitrary numbers. First of all, the equation (9b) is going to be discussed.

Left side of the equation (9b) is from matrix $\boldsymbol{B}$ and, the right side is from the matrix $\boldsymbol{\theta}$, so the matrixes are at the same kinds. so

$$\boldsymbol{B} = \sum_{n=1}^{\infty} T_n \boldsymbol{\theta}^n = \sum_{n=1}^{\infty} T_{2n} \boldsymbol{\theta}^{2n} + \sum_{n=1}^{\infty} T_{2n+1} \boldsymbol{\theta}^{2n+1} \tag{10}$$

$T_n$s are constant coefficients.

Introducing $\frac{f}{2\alpha \hbar} = \sum_{n=1}^{\infty} T_{2n} \boldsymbol{\theta}^{2n}$ and $\boldsymbol{r} = \sum_{n=1}^{\infty} T_{2n+1} \boldsymbol{\theta}^{2n+1}$,

There will be

$$\boldsymbol{B} = \frac{f}{2\alpha \hbar} + \boldsymbol{r}, \ \boldsymbol{f}^T = \boldsymbol{f}, \qquad \boldsymbol{r}^T = -\boldsymbol{r} \tag{11}$$

Where $\boldsymbol{f}$ is symmetric arbitrary matrix and $\boldsymbol{r}$ is asymmetric matrix.

By using the equations (9b) and (11), the amount of the matrix $\boldsymbol{r}$ is:

$$\boldsymbol{r} = -\frac{\boldsymbol{\theta}}{2\alpha \hbar} \tag{12}$$

Next, the equation of matrix $B$ is obtained:

$$\boldsymbol{B} = \frac{\boldsymbol{f} - \boldsymbol{\theta}}{2\alpha \hbar}, \qquad \boldsymbol{f}^T = \boldsymbol{f}, \qquad \boldsymbol{r}^T = -\boldsymbol{r} \tag{13}$$

Symmetric matrix $\boldsymbol{f}$ has indeterminate extent. This matrix is going to be discussed.

By using of equations (9c), (9d) and (13):

$$C = 0 \tag{14}$$

Thus non-commutativity equations are obtained:

$$\hat{x} = \alpha x + \left(\frac{f - \theta}{2\hbar\alpha}\right)p, \qquad \hat{p} = \beta p \tag{15}$$

Where the amount of $\alpha$ and $\beta$ numbers and matrix $f$ is indeterminated. If $=\beta = 1$, the scale of space and momentum is equaled.

$$\hat{x} = x + \left(\frac{f - \theta}{2\hbar}\right)p, \qquad \hat{p} = p \tag{16}$$

As it is determined, there is a new parameter $f$ in equations (16) where it is the same dimensionally with $\theta$.

### 4. Analysis of non-commutativity in the whole space

The whole space (momentum and space) is assumed as non-commutative in this section and by using equations (2) and (6):

$$AB^T - BA^T = AB^T - (AB^T)^T = \frac{\theta}{\hbar} \tag{17a}$$

$$CD^T - DC^T = CD^T - (CD^T)^T = \frac{\eta}{\hbar} \tag{17b}$$

$$AD^T - BC^T = I_{d\times d} \tag{17c}$$

$$\theta = -\theta^T, \qquad \eta = -\eta^T \tag{17d}$$

By investigating on equations (17a) and (17b):

$$AB^T = \frac{\theta + f_\theta}{2\hbar}, CD^T = \frac{\eta + f_\eta}{2\hbar}, f_\theta^T = f_\theta, \qquad f_\eta^T = f_\eta \tag{18}$$

There will be nothing new if the amount of $A$ and $D$ are not considered as unit matrixes because these parameters can be like a coordinate conversion. Hence, the matrixes are considered as unit matrixes ($A = D = I$). so

$$B = \frac{f_\theta - \theta}{2\hbar}, C = \frac{f_\eta + \eta}{2\hbar}, \qquad A = D = I \tag{19a}$$

$$BC^T = (f_\theta - \theta)(f_\eta - \eta) = f_\theta f_\eta - f_\theta \eta - \theta f_\eta + \theta \eta = 0 \tag{19b}$$

**Tip:** The equation (19b) demonstrates that the SW map cannot be used because it causes $\theta\eta = 0$. It means that at least, one of the non-commutative parameters is must be zero. However, the non-commutativity in the whole space (space-momentum) is the aim to be determined.

By using the equations (6) and (19), the matrixes of space and momentum is obtained:

$$\hat{x} = x + \left(\frac{f_\theta - \theta}{2\hbar}\right)p, \qquad \hat{p} = p + \left(\frac{f_\eta + \eta}{2\hbar}\right)x \tag{20}$$

By using the equation(20), commutative equation related to non-commutative equation is obtained:

$$x = \left[I - \left(\frac{f_\theta - \theta}{2\hbar}\right)\left(\frac{f_\eta + \eta}{2\hbar}\right)\right]^{-1}\left[\hat{x} - \left(\frac{f_\eta + \eta}{2\hbar}\right)\hat{p}\right], \qquad p = \left[I - \left(\frac{f_\eta + \eta}{2\hbar}\right)\left(\frac{f_\theta - \theta}{2\hbar}\right)\right]^{-1}\left[\hat{p} - \left(\frac{f_\eta + \eta}{2\hbar}\right)\hat{x}\right] \tag{21}$$

**Tip:** According to the equation (19b), If matrixes $f_\theta$ and $f_\eta$ had no amount, there should not be such an equation with SW mapping. Because their absence cause $\theta\eta = 0$ which means that one of the non-commutative parameters of space and momentum is zero.

**Tip:** As it is demonstrated, the equation (19b) makes a condition for all the equations to be confirmed. This condition will be discussed in two and three dimensions.

### 4.1. Investigation of non-commutative condition in two dimensions

The condition which was obtained in non-commutative calculations $(f_\theta - \theta)(f_\eta - \eta) = 0$ is going to be discussed for two dimensions in this section. This condition is the equation (19b) which can demonstrate the correctness of the non-commutative equation in this paper. Matrixes $f_\theta, f_\eta, \theta$ and $\eta$ are described in two dimension in this way:

$$\boldsymbol{\theta} = \theta \begin{pmatrix} 0 & 1 \\ -1 & 0 \end{pmatrix}, \boldsymbol{f_\theta} = \begin{pmatrix} f_{\theta x} & f_\theta \\ f_\theta & f_{\theta y} \end{pmatrix}, \boldsymbol{f} - \boldsymbol{\theta} = \begin{pmatrix} f_{\theta x} & f_\theta - \theta \\ f_\theta + \theta & f_{\theta y} \end{pmatrix} \qquad (22a)$$

$$\boldsymbol{\eta} = \eta \begin{pmatrix} 0 & 1 \\ -1 & 0 \end{pmatrix}, \boldsymbol{f_\eta} = \begin{pmatrix} f_{\eta x} & f_\eta \\ f_\eta & f_{\eta y} \end{pmatrix}, \boldsymbol{f_\eta} - \boldsymbol{\eta} = \begin{pmatrix} f_{\eta x} & f_\eta - \eta \\ f_\eta + \eta & f_{\eta y} \end{pmatrix} \qquad (22b)$$

The parameters $f_{\theta x}, f_{\theta y}, f_\theta, f_{\eta x}, f_{\eta y}, f_\eta$ are new ones at the equation (22) which presents non-commutativie matrixes in two dimenaions. These parametrs are meaningful in Physics and is considering in the following sections.

By using the equations (19b) and (22):

$$\boldsymbol{BC^T} = \boldsymbol{CB^T} = \begin{pmatrix} f_{\theta x}f_{\eta x} + f_\theta f_\eta + \eta f_\theta - \theta f_\eta - \eta\theta & f_{\theta x}f_\eta - \eta f_{\theta x} + f_\theta f_{\eta y} - \theta f_{\eta y} \\ f_\theta f_{\eta x} + \theta f_{\eta x} + f_\eta f_{\theta y} + \eta f_{\theta y} & f_{\theta y}f_{\eta y} + f_\theta f_\eta - \eta f_\theta + \theta f_\eta - \eta\theta \end{pmatrix} = 0 \quad (23)$$

So,

$$\begin{cases} f_{\theta x}f_{\eta x} + f_\theta f_\eta + \eta f_\theta - \theta f_\eta = \eta\theta \\ f_{\theta y}f_{\eta y} + f_\theta f_\eta - \eta f_\theta + \theta f_\eta = \eta\theta \\ f_{\theta x}f_\eta - \eta f_{\theta x} + f_\theta f_{\eta y} - \theta f_{\eta y} = 0 \\ f_\theta f_{\eta x} + \theta f_{\eta x} + f_\eta f_{\theta y} + \eta f_{\theta y} = 0 \end{cases} \qquad (24)$$

There are four equations and eight unknowns ($f_{\theta x}, f_{\theta y}, f_{\eta x}, f_{\eta y}, f_\theta, f_\eta, \theta, \eta$) in equation(24), but if we consider more in the equation, it is clear there are three equations instead of four. So there are three equations with eight unknowns. Two of the equations will be deleted by these three equations (the unknowns $f_{\theta y}, f_{\eta x}, f_{\eta y}$ are represented in terms of $f_{\theta x}, f_\theta, f_\eta, \theta, \eta$) and six unknowns are achieved. By considering equation (24):

$$f_{\theta x}f_{\eta x} = \eta\theta - f_\theta f_\eta + \mu = -(f_\eta + \eta)(f_\theta - \theta), \qquad f_{\theta y}f_{\eta y} = \eta\theta - f_\theta f_\eta - \mu = -(f_\eta - \eta)(f_\theta + \theta),$$
$$\mu = \theta f_\eta - \eta f_\theta, \qquad (25a)$$

$$f_{\eta y} = -\left(\frac{f_\eta - \eta}{f_\theta - \theta}\right) f_{\theta x}, f_{\eta x} = -\left(\frac{f_\eta + \eta}{f_\theta + \theta}\right) f_{\theta y} = -(f_\theta - \theta)(f_\eta + \eta)\frac{1}{f_{\theta x}}, \qquad (25b)$$

$$f_{\theta y} = \frac{(f_\theta + \theta)(f_\theta - \theta)}{f_{\theta x}}, \qquad f_{\eta y} = \frac{(f_\eta + \eta)(f_\eta - \eta)}{f_{\eta x}} \qquad (25c)$$

if the parameter $f_\theta$ is assumed as real as $\theta$:

$$f_{\theta y} = \frac{f_\theta^2 - \theta^2}{f_{\theta x}}, \qquad f_{\eta y} = \frac{f_\eta^2 - \eta^2}{f_{\eta x}} \qquad (25c - a)$$

If the parameter $f_\theta$ ($f_\theta^2 = f_\theta^* f_\theta \neq f_\theta f_\theta$) considers imaginative:

$$f_{\theta y} = \frac{f_\theta f_\theta - \theta^2}{f_{\theta x}}, \qquad f_{\eta y} = \frac{f_\eta f_\eta - \eta^2}{f_{\eta x}} \qquad (25c-b)$$

It will be continued by revising non-commutative space and momentum. So:

$$\boldsymbol{\theta} = \theta \begin{pmatrix} 0 & 1 \\ -1 & 0 \end{pmatrix}, \boldsymbol{f_\theta} = \begin{pmatrix} f_{\theta x} & f_\theta \\ f_\theta & f_{\theta y} \end{pmatrix}, \boldsymbol{f} - \boldsymbol{\theta} = \begin{pmatrix} f_{\theta x} & f_\theta - \theta \\ f_\theta + \theta & \frac{(f_\theta^2 - \theta^2)}{f_{\theta x}} \end{pmatrix} \qquad (26a)$$

$$\boldsymbol{\eta} = \eta \begin{pmatrix} 0 & 1 \\ -1 & 0 \end{pmatrix}, \quad \boldsymbol{f_\eta} = \begin{pmatrix} -\left(\frac{f_\eta + \eta}{f_\theta + \theta}\right) f_{\theta y} & f_\eta \\ f_\eta & -\left(\frac{f_\eta - \eta}{f_\theta - \theta}\right) f_{\theta x} \end{pmatrix},$$

$$\boldsymbol{f_\eta} + \boldsymbol{\eta} = \begin{pmatrix} -\left(\frac{f_\eta + \eta}{f_\theta + \theta}\right) f_{\theta y} & f_\eta + \eta \\ f_\eta - \eta & -\left(\frac{f_\eta - \eta}{f_\theta - \theta}\right) f_{\theta x} \end{pmatrix} \qquad (26b)$$

And,

$$\begin{pmatrix} \hat{x} \\ \hat{y} \end{pmatrix} = \begin{pmatrix} x \\ y \end{pmatrix} + \frac{1}{2\hbar} \begin{pmatrix} f_{\theta x} & f_\theta - \theta \\ f_\theta + \theta & f_{\theta y} \end{pmatrix} \begin{pmatrix} p_x \\ p_y \end{pmatrix} = \begin{pmatrix} x + \left(\frac{f_{\theta x}}{2\hbar}\right) p_x + \left(\frac{f_\theta - \theta}{2\hbar}\right) p_y \\ y + \left(\frac{f_\theta + \theta}{2\hbar}\right) p_x + \left(\frac{f_{\theta y}}{2\hbar}\right) p_y \end{pmatrix} \qquad (27a)$$

$$\begin{pmatrix} \hat{p}_x \\ \hat{p}_y \end{pmatrix} = \begin{pmatrix} p_x \\ p_y \end{pmatrix} + \frac{1}{2\hbar} \begin{pmatrix} -\left(\frac{f_\eta + \eta}{f_\theta + \theta}\right) f_{\theta y} & f_\eta + \eta \\ f_\eta - \eta & -\left(\frac{f_\eta - \eta}{f_\theta - \theta}\right) f_{\theta x} \end{pmatrix} \begin{pmatrix} x \\ y \end{pmatrix}$$

$$= \begin{pmatrix} p_x - \left(\frac{f_\eta + \eta}{f_{\theta x}}\right) \left(\frac{f_\theta - \theta}{2\hbar}\right) x + \left(\frac{f_\eta + \eta}{2\hbar}\right) y \\ p_y + \left(\frac{f_\eta - \eta}{2\hbar}\right) x - \left(\frac{f_\eta - \eta}{f_\theta - \theta}\right) \left(\frac{f_{\theta x}}{2\hbar}\right) y \end{pmatrix} \qquad (27b)$$

three dimensions is going to be discussed in some parts.

Singularity points in equation (24) are $f_\theta = \pm \theta$ and, $f_\eta = \pm \eta$ and using some of them, make non-commmutative matrix simple and other parameters zero. The first phrase causes $f_{\theta y} = f_{\theta x} = 0$ and, the second one causes $f_{\eta y} = f_{\eta x} = 0$ to become zero.

### 4.2. Analysis of non-commutative conditions in three dimensions

In this part, the condition of non commutativity is going to be discussed like the last section. First of all, it will be started with the description of non-commutative matrixes in general.

$$\boldsymbol{\theta} = \begin{pmatrix} 0 & \theta_1 & \theta_2 \\ -\theta_1 & 0 & \theta_3 \\ -\theta_2 & -\theta_3 & 0 \end{pmatrix}, \quad \boldsymbol{f_\theta} = \begin{pmatrix} f_{\theta x} & f_{\theta 1} & f_{\theta 2} \\ f_{\theta 1} & f_{\theta y} & f_{\theta 3} \\ f_{\theta 2} & f_{\theta 3} & f_{\theta z} \end{pmatrix}, \quad \boldsymbol{f_\theta} - \boldsymbol{\theta} = \begin{pmatrix} f_{\theta x} & f_{\theta 1} - \theta_1 & f_{\theta 2} - \theta_2 \\ f_{\theta 1} + \theta_1 & f_{\theta y} & f_{\theta 3} - \theta_3 \\ f_{\theta 2} + \theta_2 & f_{\theta 3} + \theta_3 & f_{\theta z} \end{pmatrix} \quad (28a)$$

$$\boldsymbol{\eta} = \begin{pmatrix} 0 & \eta_1 & \eta_2 \\ -\eta_1 & 0 & \eta_3 \\ -\eta_2 & -\eta_3 & 0 \end{pmatrix}, \quad \boldsymbol{f_\eta} = \begin{pmatrix} f_{\eta x} & f_{\eta 1} & f_{\eta 2} \\ f_{\eta 1} & f_{\eta y} & f_{\eta 3} \\ f_{\eta 2} & f_{\eta 3} & f_{\eta z} \end{pmatrix}, \quad \boldsymbol{f_\eta} - \boldsymbol{\eta} = \begin{pmatrix} f_{\eta x} & f_{\eta 1} - \eta_1 & f_{\eta 2} - \eta_2 \\ f_{\eta 1} + \eta_1 & f_{\eta y} & f_{\eta 3} - \eta_3 \\ f_{\eta 2} + \eta_2 & f_{\eta 3} + \eta_3 & f_{\eta z} \end{pmatrix} \quad (28b)$$

Here the non-commutative condition is going to be discussed, by using $(f_\theta - \theta)(f_\eta - \eta) = 0$

$$\begin{cases} f_{\theta x}f_{\eta x} + (f_\theta f_\eta)_{12} - \mu_1 - \mu_2 = (\theta\eta)_{12} \\ f_{\theta y}f_{\eta y} + (f_\theta f_\eta)_{13} + \mu_1 - \mu_3 = (\theta\eta)_{13} \\ f_{\theta z}f_{\eta z} + (f_\theta f_\eta)_{23} + \mu_2 + \mu_3 = (\theta\eta)_{23} \\ f_{\theta x}f_{\eta 1} + f_{\eta y}f_{\theta 1} - (\theta_1 f_{\eta y} + \eta_1 f_{\theta x}) - \vartheta_{23} + f_{\theta 2}f_{\eta 3} = \theta_2\eta_3 \\ f_{\theta x}f_{\eta 2} + f_{\eta z}f_{\theta 2} - (\theta_2 f_{\eta z} + \eta_2 f_{\theta x}) - \vartheta_{13} + f_{\theta 1}f_{\eta 3} = -\theta_1\eta_3 \\ f_{\eta x}f_{\theta 1} + f_{\theta y}f_{\eta 1} + (\theta_1 f_{\eta x} + \eta_1 f_{\theta y}) - \vartheta_{32} + f_{\theta 3}f_{\eta 2} = \theta_3\eta_2 \\ f_{\theta y}f_{\eta 3} + f_{\eta z}f_{\theta 3} - (\theta_3 f_{\eta z} + \eta_3 f_{\theta y}) + \vartheta_{12} + f_{\theta 1}f_{\eta 2} = \theta_1\eta_2 \\ f_{\eta x}f_{\theta 2} + f_{\theta z}f_{\eta 2} + (\theta_2 f_{\eta x} + \eta_2 f_{\theta z}) + \vartheta_{31} + f_{\theta 3}f_{\eta 1} = -\theta_3\eta_1 \\ f_{\eta y}f_{\theta 3} + f_{\theta z}f_{\eta 3} + (\theta_3 f_{\eta y} + \eta_3 f_{\theta z}) + \vartheta_{21} + f_{\theta 2}f_{\eta 1} = \theta_2\eta_1 \end{cases} \quad (29)$$

These equations below were used in the equation above:

$$\mu_1 = \theta_1 f_{\eta 1} - \eta_1 f_{\theta 1}, \quad \mu_2 = \theta_2 f_{\eta 2} - \eta_2 f_{\theta 2}, \quad \mu_3 = \theta_3 f_{\eta 3} - \eta_3 f_{\theta 3}$$

$$\mu_{12} = \mu_1 + \mu_2, \quad \mu_{13} = \mu_3 - \mu_1, \quad \mu_{23} = -\mu_3 - \mu_2$$

$$\vartheta_{12} = \theta_1 f_{\eta 2} - \eta_2 f_{\theta 1}, \quad \vartheta_{13} = \theta_1 f_{\eta 3} + \eta_3 f_{\theta 1}, \quad \vartheta_{23} = \theta_2 f_{\eta 3} - \eta_3 f_{\theta 2}$$

$$\vartheta_{21} = \theta_2 f_{\eta 1} - \eta_1 f_{\theta 2}, \quad \vartheta_{31} = \theta_3 f_{\eta 1} + \eta_1 f_{\theta 3}, \quad \vartheta_{32} = \theta_3 f_{\eta 2} - \eta_2 f_{\theta 3}$$

$$(\theta\eta)_{12} = \theta_1\eta_1 + \theta_2\eta_2, \quad (\theta\eta)_{13} = \theta_1\eta_1 + \theta_3\eta_3, \quad (\theta\eta)_{23} = \theta_2\eta_2 + \theta_3\eta_3$$

$$(f_\theta f_\eta)_{12} = f_{\theta 1}f_{\eta 1} + f_{\theta 2}f_{\eta 2}, \quad (f_\theta f_\eta)_{13} = f_{\theta 1}f_{\eta 1} + f_{\theta 3}f_{\eta 3}, \quad (f_\theta f_\eta)_{23} = f_{\theta 2}f_{\eta 2} + f_{\theta 3}f_{\eta 3}$$

There are nine equations and eighteen unknowns in equation(29). By considering equation(29):

$$f_{\theta x}f_{\eta x} = w'_{12}, \quad f_{\theta y}f_{\eta y} = w'_{13}, \quad f_{\theta z}f_{\eta z} = w'_{23}$$

$$f_{\theta y}(f_{\eta 3} - \eta_3) + f_{\eta z}(f_{\theta 3} - \theta_3) = w_{12}, \quad f_{\theta z}(f_{\eta 3} + \eta_3) + f_{\eta y}(f_{\theta 3} + \theta_3) = w_{21}$$

$$f_{\theta x}(f_{\eta 2} - \eta_2) + f_{\eta z}(f_{\theta 2} - \theta_2) = w_{13}, \quad f_{\theta z}(f_{\eta 2} + \eta_2) + f_{\eta x}(f_{\theta 2} + \theta_2) = w_{31}$$

$$f_{\theta x}(f_{\eta 1} - \eta_1) + f_{\eta y}(f_{\theta 1} - \theta_1) = w_{23}, \quad f_{\theta y}(f_{\eta 1} + \eta_1) + f_{\eta x}(f_{\theta 1} + \theta_1) = w_{32}$$

$$(30)$$

Which the equations below were used in the equation (30):

$$w_{23} = \theta_2\eta_3 + \vartheta_{23} - f_{\theta 2}f_{\eta 3}, \quad w_{32} = \theta_3\eta_2 + \vartheta_{32} - f_{\theta 3}f_{\eta 2}$$

$$w_{13} = -\theta_1\eta_3 + \vartheta_{13} - f_{\theta 1}f_{\eta 3}, \quad w_{31} = -\theta_3\eta_1 - \vartheta_{31} - f_{\theta 3}f_{\eta 1}$$

$$w_{12} = \theta_1\eta_2 - \vartheta_{12} - f_{\theta 1}f_{\eta 2}, \quad w_{21} = \theta_2\eta_1 - \vartheta_{21} - f_{\theta 2}f_{\eta 1}$$

$$w'_{12} = (\theta\eta)_{12} - (f_\theta f_\eta)_{12} + \mu_{12}, w'_{13} = (\theta\eta)_{13} - (f_\theta f_\eta)_{13} + \mu_{13}, w'_{23} = (\theta\eta)_{23} - (f_\theta f_\eta)_{12} + \mu_{12}$$

It is concluded from equation (30):

$$f_{\eta x} = \frac{w_{32} - (f_{\eta 1} + \eta_1)f_{\theta y}}{(f_{\theta 1} + \theta_1)} = \frac{w_{31} - (f_{\eta 2} + \eta_2)f_{\theta z}}{(f_{\theta 2} + \theta_2)} = \frac{w'_{12}}{f_{\theta x}}$$

$$f_{\eta y} = \frac{w_{23} - (f_{\eta 1} - \eta_1)f_{\theta x}}{(f_{\theta 1} - \theta_1)} = \frac{w_{21} - (f_{\eta 3} + \eta_3)f_{\theta z}}{(f_{\theta 3} + \theta_3)} = \frac{w'_{13}}{f_{\theta y}}$$

$$f_{\eta z} = \frac{w_{13} - (f_{\eta 2} - \eta_2)f_{\theta x}}{(f_{\theta 2} - \theta_2)} = \frac{w_{12} - (f_{\eta 3} - \eta_3)f_{\theta y}}{(f_{\theta 3} - \theta_3)} = \frac{w'_{23}}{f_{\theta z}}$$

(31)

There are nine equations and thirty unknowns in the equation(31) which five of them will be deleted by considering it $(f_{\theta y}, f_{\theta z}, f_{\eta x}, f_{\eta y}, f_{\eta z})$ and remain the other twenty five.

Since the calculations are longer in three dimensions than two, rest of the calculations are going to be passed. In the following, a free particle with a magnetic disorder in the non-commutative space is going to be discussed.

## 5. Consideration of new non-commutative equations in two dimensions for free particle equation with magnetic disorder

Hamiltonian of free particle in two dimensions for non-commutative space is going to be studied. Hamiltonian of free particle in non-commutative space is:

$$\hat{H} = \frac{\hat{p}^2}{2\hat{m}_p} \qquad (32)$$

Non-commutative matrixes (non-commutative space and momentum) in two dimensions are:

$$\begin{pmatrix} \hat{x} \\ \hat{y} \end{pmatrix} = \begin{pmatrix} x + \left(\frac{f_{\theta x}}{2\hbar}\right)p_x + \left(\frac{f_\theta - \theta}{2\hbar}\right)p_y \\ y + \left(\frac{f_\theta + \theta}{2\hbar}\right)p_x + \left(\frac{f_{\theta y}}{2\hbar}\right)p_y \end{pmatrix} \qquad (33a)$$

$$\begin{pmatrix} \hat{p}_x \\ \hat{p}_y \end{pmatrix} = \begin{pmatrix} p_x - \left(\frac{f_\eta + \eta}{f_\theta + \theta}\right)\left(\frac{f_{\theta y}}{2\hbar}\right)x + \left(\frac{f_\eta + \eta}{2\hbar}\right)y \\ p_y + \left(\frac{f_\eta - \eta}{2\hbar}\right)x - \left(\frac{f_\eta - \eta}{f_\theta - \theta}\right)\left(\frac{f_{\theta x}}{2\hbar}\right)y \end{pmatrix} \qquad (33b)$$

It is assumed that $f_{\theta x}$ and $f_{\theta y}$ are real parameters in this part (this assumption is just in this case. The imaginary part can be added in other cases, though. Hence we suggest to consider $f_{\theta x}$ and $f_{\theta y}$ for imaginary part). Non-commutative equation is pasted in the equation of free particle's Hamiltonian ($\hat{H} = \frac{\hat{p}^2}{2\hat{m}_p}$), then:

$$\hat{H} = \frac{\left[p_x - \left(\frac{f_\eta + \eta}{f_\theta + \theta}\right)\left(\frac{f_{\theta y}}{2\hbar}\right)x + \left(\frac{f_\eta + \eta}{2\hbar}\right)y\right]^2}{2\hat{m}_p} + \frac{\left[p_y + \left(\frac{f_\eta - \eta}{2\hbar}\right)x - \left(\frac{f_\eta - \eta}{f_\theta - \theta}\right)\left(\frac{f_{\theta x}}{2\hbar}\right)y\right]^2}{2\hat{m}_p} \qquad (34)$$

**Tip:** $f_\theta$ and $f_\eta$ are real parameters.

Hamiltonian of equation (34) is monotonic in magnetic field as the same as the Hamiltonian of free particle. So the Hamiltonian of free particle in magnetic field is going to be discussed then coefficients of them will be compared.

Potential of magnetic field is considered as:

$$A_x = \alpha_x x + \beta_x y, \qquad A_y = \alpha_y x + \beta_y y, \qquad B_z = (\alpha_y - \beta_x) \qquad (35)$$

So,

$$H = \frac{\left(p_x - \frac{e}{c}(\alpha_x x + \beta_x y)\right)^2 + \left(p_y - \frac{e}{c}(\alpha_y x + \beta_y y)\right)^2}{2m_p} \tag{36}$$

Comparison of the equations (34) and (36) and assumption of <u>equaled-mass</u> in non-commutative space and commutative space ($\hat{m}_p = m_p$), it is obtained:

$$\left(\frac{f_\eta + \eta}{f_\theta + \theta}\right)\left(\frac{f_{\theta y}}{2\hbar}\right) = \frac{e}{c}\alpha_x, \quad -\left(\frac{f_\eta + \eta}{2\hbar}\right) = \frac{e}{c}\beta_x, \quad -\left(\frac{f_\eta - \eta}{2\hbar}\right) = \frac{e}{c}\alpha_y, \quad \left(\frac{f_\eta - \eta}{f_\theta - \theta}\right)\left(\frac{f_{\theta x}}{2\hbar}\right) = \frac{e}{c}\beta_y \tag{37}$$

Non-commutative momentum has been equaled with disordered momentum in the equation above. By simplifying the equation(37):

$$f_\eta = -\frac{2\hbar e}{c}\beta_x - \eta = -\frac{2\hbar e}{c}\alpha_y + \eta = -\frac{\hbar e}{c}(\beta_x + \alpha_y), \quad \eta = \frac{\hbar e}{c}(\alpha_y - \beta_x) = \frac{\hbar e}{c}B_z \tag{38a}$$

$$f_{\theta x} = -\frac{\alpha_x}{\beta_x}(f_\theta - \theta), \quad f_{\theta y} = -\frac{\beta_y}{\alpha_y}(f_\theta + \theta) \tag{38b}$$

According to equation(25c), it is obtained:

$$\frac{\beta_y}{\beta_x} = \frac{\alpha_y}{\alpha_x}, \quad f_{\theta x} = -\frac{\alpha_x}{\beta_x}(f_\theta - \theta), \quad f_{\theta y} = -\frac{\beta_x}{\alpha_x}(f_\theta + \theta) \tag{38c}$$

The importance of $f_\eta$ is obvious in equation(38a). Its nature is a magnetic field. Equation (38c) demonstrates the corresponding of parameters $f_\theta$, $f_{\theta x}$ and $f_{\theta y}$ to $\theta$ and each other. Also this equation indicates one equation with three unknowns for non-commutative parameters. So always there are two independent parameters.

At last, the parameters momentum, space and Hamiltonian are:

$$\begin{pmatrix}\hat{x}\\\hat{y}\end{pmatrix} = \begin{pmatrix}x - \frac{\alpha_x}{\beta_x}\left(\frac{f_\theta - \theta}{2\hbar}\right)p_x + \left(\frac{f_\theta - \theta}{2\hbar}\right)p_y\\ y + \left(\frac{f_\theta + \theta}{2\hbar}\right)p_x - \frac{\beta_x}{\alpha_x}\left(\frac{f_\theta + \theta}{2\hbar}\right)p_y\end{pmatrix} \tag{33a}$$

$$\begin{pmatrix}\hat{p}_x\\\hat{p}_y\end{pmatrix} = \begin{pmatrix}p_x - \frac{e}{c}(\alpha_x x + \beta_x y)\\ p_y - \frac{e}{c}(\alpha_y x + \beta_y y)\end{pmatrix} \tag{39b}$$

Here it is found out that non-commutative equations can affect the free particle in magnetic field easily. All is needed to calculate it, is to equalize the non-commutativity and magnetic disorder. The effect of non-commutativity on motion equations are going to be discussed in the following.

Some important matters are expressed for better understanding:

1. The space of non-disordered is as the same as the space of disordered.
2. Frequency is changed if a disorder is created in the commutative space. However the frequency of commutative space is the same as the frequency of non-commutative space.
3. Function mode of disordered commutative space is as the same as the function mode of disordered non-commutative space.
4. By using Heisenberg motion equation ($\frac{d\langle A\rangle}{dt} = \langle\frac{\partial A}{\partial t}\rangle + \frac{1}{i\hbar}\langle[A,H]\rangle$) motion equation of particle in non-commutative space is going to be discussed. it should be noticed that if state function is described as non-disordered commutative ($\psi = \psi(x_i, t)$), then $\langle\frac{\partial x_i}{\partial t}\rangle = \langle\frac{\partial p_i}{\partial t}\rangle = 0$. Also, if the non-commutative parameters are dependent on time therefore, $\langle\frac{\partial \hat{x}_i}{\partial t}\rangle \neq 0, \langle\frac{\partial \hat{p}_i}{\partial t}\rangle \neq 0$. But if the state function is described as non-commutative ($\psi = \psi(\hat{x}_i, t)$) therefore $\langle\frac{\partial \hat{x}_i}{\partial t}\rangle = \langle\frac{\partial \hat{p}_i}{\partial t}\rangle = 0$. Also, non-disordered commutative parameters $\langle\frac{\partial x_i}{\partial t}\rangle, \langle\frac{\partial p_i}{\partial t}\rangle$ can be opposite of zero but, disordered commutative parameters are zero ($\langle\frac{\partial x_i}{\partial t}\rangle_{Disorder} = \langle\frac{\partial p_i}{\partial t}\rangle_{Disorder} = 0$). Since disordered commutative equations are equalizing with non-disordered non-commutative ones, state function is in non-commutative space or disordered commutativity. It means that Heisenberg motion equation is confirmed $\frac{d\langle A\rangle}{dt} = \frac{1}{i\hbar}\langle[A, H]\rangle$ for both of them.

5. In the following calculations, this symbol $\langle\ \rangle$ is going to be skipped ($\frac{dA}{dt} = \frac{\partial A}{\partial t} + \frac{1}{i\hbar}[A, H]$).

Now motion equations are going to be discussed:

$$\frac{dx}{dt} = \frac{1}{i\hbar}\left[x, \frac{\left(p_x - \frac{e}{c}(\alpha_x x + \beta_x y)\right)^2 + \left(p_y - \frac{e}{c}(\alpha_y x + \beta_y y)\right)^2}{2m_p}\right] \quad (40a)$$

$$\frac{dy}{dt} = \frac{1}{i\hbar}\left[y, \frac{\left(p_x - \frac{e}{c}(\alpha_x x + \beta_x y)\right)^2 + \left(p_y - \frac{e}{c}(\alpha_y x + \beta_y y)\right)^2}{2m_p}\right] \quad (40b)$$

$$\frac{dp_x}{dt} = \frac{1}{i\hbar}\left[p_x, \frac{\left(p_x - \frac{e}{c}(\alpha_x x + \beta_x y)\right)^2 + \left(p_y - \frac{e}{c}(\alpha_y x + \beta_y y)\right)^2}{2m_p}\right] \quad (40c)$$

These equations are obtained by simplifying the equations (40)

$$\frac{dx}{dt} = \frac{1}{m_p}p_x - \frac{e}{cm_p}(\alpha_x x + \beta_x y) \quad (41a)$$

$$\frac{dy}{dt} = \frac{1}{m_p}p_y - \frac{e}{cm_p}(\alpha_y x + \beta_y y) \quad (41b)$$

$$\frac{dp_x}{dt} = \frac{e}{cm_p}\left[\alpha_x p_x + \alpha_y p_y - \frac{e}{c}[(\alpha_x^2 + \alpha_y^2)x + (\alpha_x \beta_x + \alpha_y \beta_y)y]\right] \quad (41c)$$

$$\frac{dp_y}{dt} = \frac{e}{cm_p}\left[\beta_x p_x + \beta_y p_y - \frac{e}{c}[(\alpha_x \beta_x + \alpha_y \beta_y)x + (\beta_x^2 + \beta_y^2)y]\right] \quad (41d)$$

Differential equations of all space and momentum parameters can be achieved separately by using equations (41).

$$\frac{d^2x}{dt^2} + \omega\frac{dy}{dt} = 0, \qquad \frac{d^2y}{dt^2} - \omega\frac{dx}{dt} = 0 \quad (42a)$$

$$p_x = m_p\frac{dx}{dt} + \frac{e}{c}(\alpha_x x + \beta_x y), \qquad p_y = m_p\frac{dy}{dt} + \frac{e}{c}(\alpha_y x + \beta_y y) \quad (42b)$$

Here $\omega = \frac{e}{cm_p}(\beta_x - \alpha_y) = -\frac{e}{cm_p}B_z$.

It is assumed in the equations above that mounts of ($\alpha_x, \alpha_y, \beta_x, \beta_y$) are constant (for simplicity in calculation) and, do not change over time. If these amounts are not assumed constantly commutative amounts will be variable as well and, equation (42a) will change. Space equations can be achieved by using of equation (42a).

$$x = x_3 + x_1\sin(\omega t) + x_2\cos(\omega t), \qquad y = y_3 + y_1\sin(\omega t) + y_2\cos(\omega t) \quad (43)$$

By using of equations (42a) and, (43) :

$$y_2 = -x_1, \quad y_1 = x_2 \tag{44}$$

Therefore,

$$x = x_3 + x_1 \sin(\omega t) + x_2 \cos(\omega t), \quad y = y_3 + x_2 \sin(\omega t) - x_1 \cos(\omega t), \quad \omega = -\frac{e}{cm_p} B_z \tag{45}$$

By using equations (42b) and, (45) :

$$p_x = \frac{e}{c}(\alpha_x x_1 + \alpha_y x_2) \sin(\omega t) + \frac{e}{c}(\alpha_x x_2 - \alpha_y x_1) \cos(\omega t) + \frac{e}{c}(\alpha_x x_3 + \beta_x y_3) \tag{46a}$$

$$p_y = \frac{e}{c}(\beta_y x_2 + \beta_x x_1) \sin(\omega t) + \frac{e}{c}(\beta_x x_2 - \beta_y x_1) \cos(\omega t) + \frac{e}{c}(\alpha_y x_3 + \beta_y y_3) \tag{46b}$$

Then, all previous calculations are going to be calculated for non-commutative space and momentum:

$$\frac{d\hat{x}}{dt} = \frac{1}{i\hbar}\left[\hat{x}, \frac{\hat{p}_x^2 + \hat{p}_y^2}{2m_p}\right] \tag{47a}$$

$$\frac{d\hat{y}}{dt} = \frac{1}{i\hbar}\left[\hat{y}, \frac{\hat{p}_x^2 + \hat{p}_y^2}{2m_p}\right] \tag{47b}$$

$$\frac{d\hat{p}_x}{dt} = \frac{1}{i\hbar}\left[\hat{p}_x, \frac{\hat{p}_x^2 + \hat{p}_y^2}{2m_p}\right] \tag{47c}$$

$$\frac{d\hat{p}_y}{dt} = \frac{1}{i\hbar}\left[\hat{p}_y, \frac{\hat{p}_x^2 + \hat{p}_y^2}{2m_p}\right] \tag{47d}$$

Before considering equations above, equations below are remaindered.

$$[\hat{x}_i, \hat{p}_j^2] = \hat{x}_i \hat{p}_j^2 - \hat{p}_j^2 \hat{x}_i + \hat{p}_j \hat{x}_i \hat{p}_j - \hat{p}_j \hat{x}_i \hat{p}_j = [\hat{x}_i, \hat{p}_j]\hat{p}_j + \hat{p}_j[\hat{x}_i, \hat{p}_j] = 2i\hbar \delta_{ij}\hat{p}_j$$

$$[\hat{p}_i, \hat{p}_j^2] = \hat{p}_i \hat{p}_j^2 - \hat{p}_j^2 \hat{p}_i + \hat{p}_j \hat{p}_i \hat{p}_j - \hat{p}_j \hat{p}_i \hat{p}_j = [\hat{p}_i, \hat{p}_j]\hat{p}_j + \hat{p}_j[\hat{p}_i, \hat{p}_j] = 2i\hbar \eta_{ij}\hat{p}_j$$

The equations below are achieved by solving equations (47):

$$\frac{d\hat{x}}{dt} = \frac{\hat{p}_x}{m_p} \tag{48a}$$

$$\frac{d\hat{y}}{dt} = \frac{\hat{p}_y}{m_p} \tag{48b}$$

$$\frac{d\hat{p}_x}{dt} = -\frac{e}{cm_p}(\beta_x - \alpha_y)\hat{p}_y \tag{48c}$$

$$\frac{d\hat{p}_y}{dt} = \frac{e}{cm_p}(\beta_x - \alpha_y)\hat{p}_x \tag{48d}$$

Equations (48c) and, (48d) are easily solvable and can be turned to linear equation as below.

$$\frac{d^2\hat{p}_x}{dt^2} = -\omega^2 \hat{p}_x, \quad \frac{d^2\hat{p}_y}{dt^2} = -\omega^2 \hat{p}_y \tag{49}$$

And $\omega = \frac{e}{cm_p}(\beta_x - \alpha_y) = -\frac{e}{cm_p}B_z = \frac{2\eta}{m_p \hbar}$.

Solutions of equation (49) is:

$$\hat{p}_x = \hat{p}_{1x} \sin \omega t + \hat{p}_{2x} \cos \omega t, \hat{p}_y = \hat{p}_{1y} \sin \omega t + \hat{p}_{2y} \cos \omega t, \qquad \hat{p}_{2y} = -\hat{p}_{1x}, \qquad \hat{p}_{1y} = \hat{p}_{2x} \quad (50a)$$

$$\hat{p}_x = \hat{p}_{1x} \sin \omega t + \hat{p}_{2x} \cos \omega t, \hat{p}_y = \hat{p}_{2x} \sin \omega t - \hat{p}_{1x} \cos \omega t, \qquad \omega = -\frac{e}{cm_p}B_z \quad (50b)$$

Amount of non-commutativity are equaled with commutativity in all times for equalizing constant non-commutative parameters with commutative ones in order to, proportion of their commutativity with non-commutativity achieves.

By using equations (39b), (45) and, (48):

$$\hat{p}_x = -\frac{e}{c}(\beta_x - \alpha_y)x_2 \sin(\omega t) + \frac{e}{c}(\beta_x - \alpha_y)x_1 \cos(\omega t) \tag{51a}$$

$$\hat{p}_y = \frac{e}{c}(\beta_x - \alpha_y)x_1 \sin(\omega t) + \frac{e}{c}(\beta_x - \alpha_y)x_2 \cos(\omega t) \tag{51b}$$

By using equations (50) and, (51)

$$\hat{p}_{1x} = -\frac{e}{c}(\beta_x - \alpha_y)x_2 = \frac{e}{c}B_z x_2, \qquad \hat{p}_{2x} = \frac{e}{c}(\beta_x - \alpha_y)x_1 = -\frac{e}{c}B_z x_1 \tag{52}$$

Thus, equation of momentum in non-commutative space is achieved as below:

$$\hat{p}_x = \frac{e}{c}B_z x_2 \sin \omega t - \frac{e}{c}B_z x_1 \cos \omega t, \qquad \hat{p}_y = -\frac{e}{c}B_z x_1 \sin \omega t - \frac{e}{c}B_z x_2 \cos \omega t \tag{53}$$

By considering equations (45) and (53), It is achieved that $\hat{p}_x = m_p \frac{dx}{dt}$ and, $\hat{p}_y = m_p \frac{dy}{dt}$. This means that there is a relation between new commutative space and momentum non-commutative.

Now, equations (48a) and, (48b) can be revised as below:

$$\frac{d\hat{x}}{dt} = \frac{dx}{dt} + \frac{1}{2\hbar}\frac{d(f_\theta - \theta)}{dt}\left(\frac{\beta_y}{\beta_x}p_x + p_y\right) + \frac{(f_\theta - \theta)}{2\hbar}\frac{d}{dt}\left(\frac{\beta_y}{\beta_x}p_x + p_y\right) = \frac{\hat{p}_x}{m_p} \tag{54a}$$

$$\frac{d\hat{y}}{dt} = \frac{dy}{dt} + \frac{1}{2\hbar}\frac{d(f_\theta + \theta)}{dt}\left(p_x + \frac{\beta_x}{\beta_y}p_y\right) + \frac{(f_\theta + \theta)}{2\hbar}\frac{d}{dt}\left(p_x + \frac{\beta_x}{\beta_y}p_y\right) = \frac{\hat{p}_y}{m_p} \tag{54b}$$

By using equations (54):

$$\frac{1}{2\hbar}\frac{d(f_\theta - \theta)}{dt}\left(\frac{\beta_y}{\beta_x}p_x + p_y\right) + \frac{(f_\theta - \theta)}{2\hbar}\frac{d}{dt}\left(\frac{\beta_y}{\beta_x}p_x + p_y\right) = 0 \tag{55a}$$

$$\frac{1}{2\hbar}\frac{d(f_\theta + \theta)}{dt}\left(p_x + \frac{\beta_x}{\beta_y}p_y\right) + \frac{(f_\theta + \theta)}{2\hbar}\frac{d}{dt}\left(p_x + \frac{\beta_x}{\beta_y}p_y\right) = 0 \tag{55b}$$

By using equations (46), equations below are achieved:

$$u = \frac{\beta_y}{\beta_x}p_x + p_y = \frac{e}{c}(\alpha_y + \beta_x)\left\{\left[\frac{\beta_y}{\beta_x}x_2 + x_1\right]\sin(\omega t) + \left[x_2 - \frac{\beta_y}{\beta_x}x_1\right]\cos(\omega t)\right\} + 2\frac{e}{c}(\alpha_y x_3 + \beta_y y_3) \tag{56a}$$

$$v = p_x + \frac{\beta_x}{\beta_y}p_y = \frac{e}{c}(\alpha_y + \beta_x)\left\{\left[x_2 + \frac{\beta_x}{\beta_y}x_1\right]\sin(\omega t) + \left[\frac{\beta_x}{\beta_y}x_2 - x_1\right]\cos(\omega t)\right\} + 2\frac{e}{c}\frac{\beta_x}{\beta_y}(\alpha_y x_3 + \beta_y y_3) \tag{56b}$$

Which here $u$ and, $v$ are new figures of momentum.

By using equations (56):

$$v = \frac{\beta_x}{\beta_y} u \tag{57}$$

By using equations (55) and, (57):

$$\frac{d(f_\theta - \theta)}{dt} u + (f_\theta - \theta)\frac{du}{dt} = 0 \tag{58a}$$

$$\frac{\beta_x}{\beta_y}\frac{d(f_\theta + \theta)}{dt} u + \frac{\beta_x}{\beta_y}(f_\theta + \theta)\frac{du}{dt} = 0 \tag{58b}$$

One of the answers of non-commutative parameters is zero ($f_\theta = \theta = 0$) but, if it is assumed that these parameters are not zero then, some approaches are applicated for considering equation (58).

**Approach One:** if $\frac{\beta_x}{\beta_y} \neq 0$ then, there is no new extent for fields. Also:

$$\frac{d(f_\theta - \theta)}{dt} u + (f_\theta - \theta)\frac{du}{dt} = 0 \tag{59a}$$

$$\frac{d(f_\theta + \theta)}{dt} u + (f_\theta + \theta)\frac{du}{dt} = 0 \tag{59b}$$

So,

$$\frac{\frac{d(f_\theta - \theta)}{dt}}{(f_\theta - \theta)} = \frac{\frac{d(f_\theta + \theta)}{dt}}{(f_\theta + \theta)} = -\frac{\frac{du}{dt}}{u} \tag{60}$$

By solving equation (60)

$$\ln c_-(f_\theta - \theta) = \ln c_+(f_\theta + \theta) = -\ln u \tag{61}$$

Which here $c_-$ and, $c_+$ are integral constants.

So,

$$f_\theta = \frac{\left(\frac{c_-}{c_+} + 1\right)}{\left(\frac{c_-}{c_+} - 1\right)} \theta = \frac{\left(\frac{c_-}{c_+} + 1\right)}{2c_-}\frac{1}{u}, \qquad \theta = \frac{\left(\frac{c_-}{c_+} - 1\right)}{2c_-}\frac{1}{u} \tag{62}$$

As it is clear from equation (62) new parameters appear then, gain physical concept (parameters $\theta$ and, $f_\theta$ own reverse dimensions of momentum and, connect to each other with a constant coefficient). This means that non-commutativity is not a simple disorder rather owns information of non-commutative space in the whole universe. Interpretation of this parameter needs more basic concepts. Because these kinds of equations were not foreseen. As it is obvious non-commutative space parameters declare their dependency on time. Here, a space is discovered equal to prior space which the result of both of them are the same. The most important thing is that the equivalent space is considerable in this case but, if the oscillator topic is raised, these parameters are needed to justify these equations.

**Approach Two:** if $\frac{\beta_x}{\beta_y} = 0$ then, $\frac{\alpha_x}{\alpha_y} = 0$ which brings the issue to some extents in amount of field so, this would not be the proper approach. However, amount of non-commutative space parameter will stand independent and equation below is confirmed for it

$$c_-(f_\theta - \theta) = \frac{1}{u} \tag{71}$$

Here, we move on discussing about non-commutativity in two-dimensional harmonic oscillator.

## 6. Conclusion

non-commutative equations were described with a magnetic field. The subject which was declared in this paper was the role of new parameters in standardization and describing fields which they are meaningful at the both views. Even in most cases, these optional parameters demonstrate new meanings such as equivalent of an infinite equivalent space which have same actions but different structures. In fact, this matter can be described as an electrical charge sea which is just an electric charge point in observer's sight but, there are so many of electrical charges practically (creation and destruction of the particle). These calculations have opened a new realization in the depth of non-commutative space which can be noticed more extensive than the past. Even, measurement of mass in the two commutative and non-commutative spaces can be mentioned and, a new kind of rotation will be discussed which its effect on the mass is measurable.

Furthermore, equation (62) demonstrates that new parameters are defined by momentum inverse. In fact, it describes a disorder which was not detected before and, can be made of electrical field. So, these parameters help the science of elementary particles to have more accurate corrections such as corrections in Nucleon Vertex functions and, Electromagnetic functions. To put differently, new parameters are hidden ones in the calculations of Uncertainty Principle and their role get important in low energies.

## 7. Suggestions

Suggestions will be introduced in the following sections:

1. Investigate the particle's calculations in magnetic field for imaginary spaces.
2. Investigate the harmonic oscillator
3. Investigate the equations in 3 dimensions and even more
4. Investigate the non-commutative space for correcting relativity
5. Investigate the non-commutative space for Dirac equation
6. Investigate the non-commutative space for Hydrogen atom equations, relativity and, …